\documentclass[
preprint,
amsmath,amssymb,
aip,
rsi,
]{revtex4-1}

\usepackage{amsmath,amssymb}
\usepackage{subfigure}
\usepackage{mdframed} 
\usepackage{graphicx}
\usepackage{hyperref}
\usepackage{bm}
\usepackage{dcolumn}
\hypersetup{
	colorlinks = true, linkcolor = blue, anchorcolor = blue, urlcolor = blue, citecolor = blue
}

\graphicspath{{./Figures2/}}

\begin{document}

\title{Scalable Cryogenic Read-out Circuit for a Superconducting Nanowire Single-Photon Detector System}

\author{Clinton Cahall}\email{clinton.cahall@duke.edu}
\affiliation{Department of Electrical and Computer Engineering, Duke University, Durham, NC 27708, USA}

\author{Daniel J. Gauthier}
\affiliation{Department of Physics, The Ohio State University, Columbus, OH 43210, USA}

\author{Jungsang Kim}
\affiliation{Department of Electrical and Computer Engineering, Duke University, Durham, NC 27708, USA}
\affiliation{IonQ, Inc., College Park, MD 20740, USA}

\date{\today}

\begin{abstract}
The superconducting nanowire single-photon detector (SNSPD) is a leading technology for quantum information science applications using photons, and they are finding increasing uses in photon-starved classical imaging applications. Critical detector characteristics, such as timing resolution (jitter), reset time and maximum count rate, are heavily influenced by the read-out electronics that sense and amplify the photon detection signal. We describe a read-out circuit for SNSPDs using commercial off-the-shelf amplifiers operating at cryogenic temperatures.  Our design demonstrates a 35\,ps timing resolution and a maximum count rate of over $2\times10^{7}$ counts per second while maintaining $<3$\,mW power consumption per channel, making it suitable for a multichannel read-out.
\end{abstract}


\maketitle

\section{Introduction}

\hspace{.01em}
\indent 
Superconducting nanowire single-photon detectors (SNSPD) are one of the leading solutions for high performance photon detection in quantum information and communication applications. SNSPDs played a crucial role in the recent loophole-free Bell inequality measurements~\cite{shalm2015strong} and the Lunar Laser Communication Demonstration.~\cite{boroson2014overview} These detectors feature high detection efficiency,~\cite{dauler2014} low dark count rates, good timing resolution and short reset times, and enhance the performance of quantum communication protocols such as quantum key distribution (QKD) and superdense quantum teleportation.~\cite{graham2015}

A superconducting nanowire photon detector was first realized in 2001 by Gol'tsman and colleagues at the University of Moscow.~\cite{semenov2001} They were able to detect optical photons with a 225-nm-wide, 1-$\mu$m-long niobium nitride (NbN) wire cooled to 4.2\,K. Since the initial demonstration of an SNSPD, several groups have advanced this technology to improve optical coupling,~\cite{miller2011compact} increase detection efficiency,~\cite{rosenberg2004near,rosfjord2006nanowire} investigate alternate superconducting materials,~\cite{baek2011superconducting,verma2014,verma2015high} and develop complex detector structures including pixel arrays~\cite{dauler2009photon,verma2014four,allman2015near} and parallel elements.~\cite{najafi2012timing,zhao2014eight} These advancements and recent demonstrations of high detection efficiency,~\cite{marsili2013} high maximum count rate~\cite{kerman2006kinetic} and good timing resolution,~\cite{esmaeil2017single} have enabled SNSPDs to be a leading commercial technology~\cite{snspd} for quantum optics experiments. 

Critical detector characteristics, such as timing jitter, reset times and maximum count rates,~\cite{kerman2013} are often dictated by the read-out electronics that sense and amplify the electrical signal generated in response to a photon detection event. In this paper, we describe the design and performance of cryogenic amplifiers that provide two critical advantages for SNSPD read-out: $(1)$ the intrinsic amplifier noise can be reduced dramatically, improving the signal-to-noise ratio (SNR) and hence the timing jitter;~\cite{wu2017improving} and $(2)$ placing the first-stage preamplifier closer to the device provides flexibility to design the effective load impedance of the amplifier with minimal signal loss between the detector and the preamplifier. There have been recent demonstrations of low-noise, low-power cryogenic amplifiers built from custom-made components~\cite{bardin2009,montazeri2016ultra} and commercially available transistors.~\cite{mani2014single} The main goal of our effort is to identify a commercial monolithic amplifier read-out circuit that minimizes timing jitter while maintaining high count rate capabilities, and scalability in terms of cost and power consumption. These characteristics have a significant impact on the speed and fidelity of quantum communication protocols such as time-bin encoded QKD.~\cite{brougham2016} Recently, we used our amplifier read-out scheme in a high-rate QKD demonstration.~\cite{islam2017robust,islam2017provably}

\section{Low Temperature Read-out Scheme}

We explore a range of read-out-circuit strategies using commercial gallium arsenide (GaAs) monolithic microwave integrated circuits (MMIC). Unlike conventional silicon devices, GaAs chips do not suffer from carrier freeze-out at cryogenic temperatures, and provide a viable solution for the SNSPD read-out circuit. We compare our read-out circuit to the high-performance cryogenic amplifier CITLF\,3 built by Cosmic Microwave Technology, formerly part of the Microwave Research Group at California Institute of Technology. The CITLF\,3 features 30\,dB gain over a bandwidth of 2.5\,GHz, a noise temperature of 4\,K, dissipates 24\,mW of power, and carries a high cost. We also compare our results to the performance of commercial low-noise room-temperature amplifiers (Miteq AU1310).

{
\begin{figure}
\includegraphics[width=1.0\linewidth]{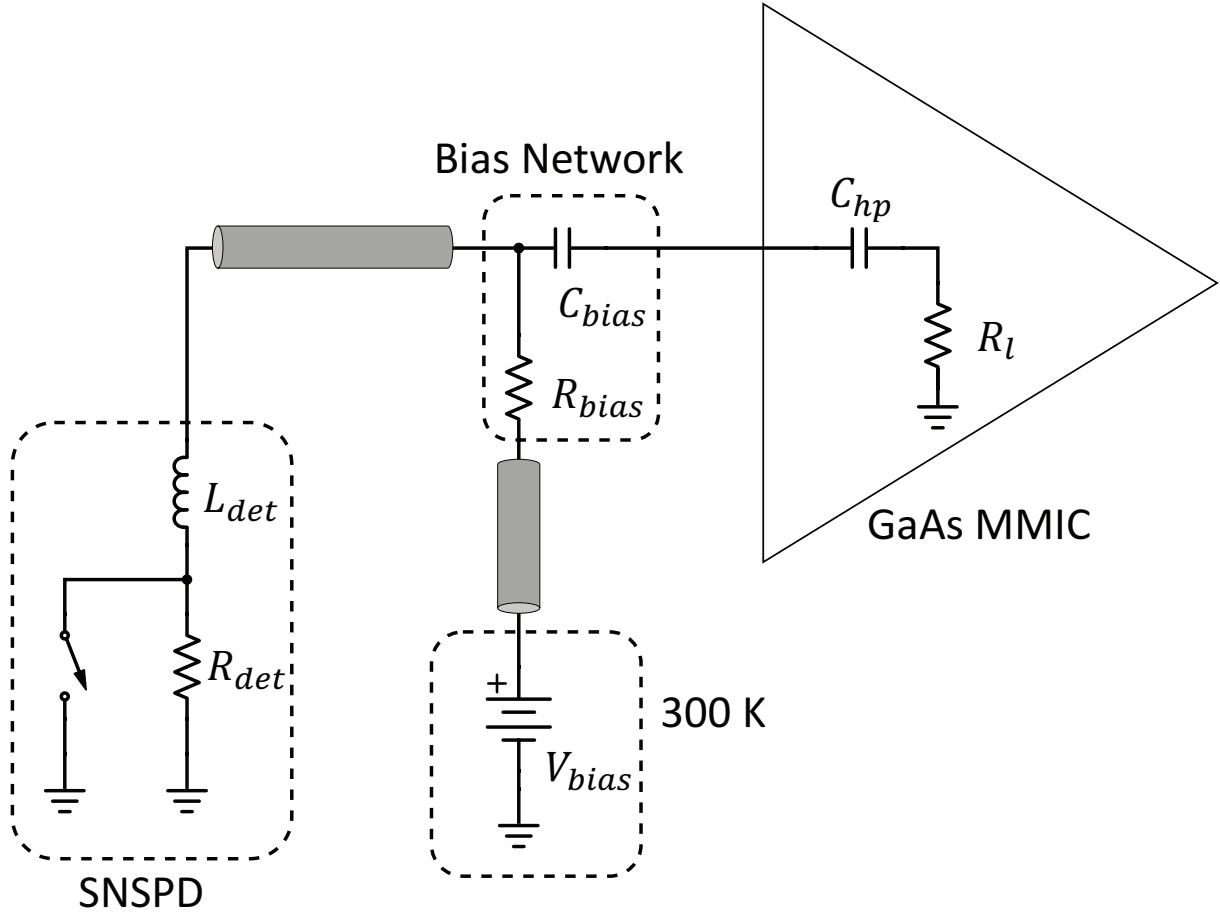}
\caption{Simplified circuit schematic of the detector and pre-amplifier setup. The bias-network on the amplifier front-end has component values $C_{bias}=100$\,pF (Vishay Vitramon, P/N: VJ0603A101KXBAC31) and $R_{bias} = 100$\,k$\Omega$ (Susumu, P/N: RR1220P-104-D). The resistor in the bias-network has a large value to act as a constant current source. The high-pass corner frequency and input impedance are set by the series capacitor $C_{hp}$ and resistor to ground $R_{l}$. These components are internal to the amplifier, and the values depend strongly on the operating temperature.}
\label{fig1}
\end{figure}
}

\indent
Our most successful circuit uses a GaAs MMIC amplifier chip from Broadcom Limited, formerly Avago Tech. (MGA-$68563$). Amplifier prototypes are assembled on patterned Rogers 4003C high frequency printed circuit board (PCB) with passive components selected for cryogenic compatibility, such as thin film resistors~\cite{lamb2014evaluation} and ceramic capacitors using C0G/NP0 dielectric.~\cite{teyssandier2010commercially} The amplifier consists of bias circuitry for powering the amplifier as well as a custom bias network used to split the DC current bias to the detector and the high-frequency detection signal at the amplifier input. A schematic of the detector and read-out circuit are shown in Fig.~\ref{fig1}.

We test our read-out circuit on single pixel SNSPDs made of a proprietary amorphous superconducting material from Quantum Opus. Our cryostat system consists of a Gifford-McMahon (GM) cryocooler and a closed-cycle helium-4 sorption refrigerator featuring a base temperature of 0.8\,K, designed to support the operation of SNSPDs made out of tungsten silicide (WSi), molybdenum silicide (MoSi), and other amorphous superconducting materials. The 4\,K-stage of the GM cryocooler provides sufficient cooling power (200\,mW at 4.2\,K), adequate for thermally anchoring the low temperature preamplifiers.

To understand our design choices for the read-out circuit, we briefly highlight the characteristics of the SNSPD used in our study. A photon absorbed in a superconducting nanowire biased with a DC current deposits enough energy in the wire to heat it locally above the superconducting critical temperature and hence to the normal, resistive state. The temperature of the hot-spot region grows due to Joule heating and the resistive section quickly spreads due to diffusion of non-equilibrium quasi-particles~\cite{semenov2001,renema2014experimental} until a section of the wire is resistive. The hot-spot resistance depends on the material parameters as well as the electrical readout circuit~\cite{kerman2009electrothermal} and it typically is in the range of 1-10\,k$\Omega$.

The SNSPD is modeled as an inductor $L_{det}$ in series with a resistor $R_{det}$ to ground, as shown in Fig.~\ref{fig1}. In the detector quiescent state, the nanowire is superconducting and the switch in parallel with the resistor is closed. A photon detection event corresponds to the switch opening, leading to the detector resistance $R_{det}$, where $R_{det}\simeq2$\,k$\Omega$ in our device and our read-out circuit. As the detector cools down and the nanowire returns to the superconducting state, the switch closes. The inductance arises from the kinetic inductance of the superconducting nanowire, typically in the range of $\sim1\,\mu$H for the devices we tested.

When a photon detection event drives the detector to a normal state, the bias current through the detector is diverted to the amplifier's input with a load resistance of $R_{l}$ provided by the input impedance of the amplifier, and this current is detected as an output pulse. The rise time of the detected pulse is determined by  $\tau_{R}\approx L_{det}/R_{det}$, and the recovery time for the pulse to fully decay and the detector to reset is determined by $\tau_{Decay}\approx L_{det}/R_{l}$.

The values for $L_{det}$ and $R_{det}$ are estimated using each of these time-constants of the detection waveform. The value for $L_{det}$ is estimated by fitting an exponential function to the recovery of the waveform back to zero using the value for the read-out circuit load resistance, such as a $50\,\Omega$ room temperature amplifier. Once $L_{det}$ is known, the value for $R_{det}$ is estimated by measuring the rise-time of a detection waveform. A high bandwidth read-out must be used in order to limit the effects that a bandwidth-limited read-out circuit has on the fast rising edge of the waveform.

A read-out circuit optimized for low timing jitter and maximum achievable count rate is low noise, has a large bandwidth, and is DC coupled at the amplifier input. Electrical read-out noise is the dominant contribution to the measured timing jitter and is approximately quantified by the rise time of the photon detection waveform divided by the SNR.~\cite{marsili2013} Timing performance is improved by a reduction in the amplifier noise and securing a large bandwidth to improve the rise time of the photon detection waveform. The relaxation time of a photon hot-spot is on the order of $\sim500$\,ps~\cite{marsili2016hotspot} and so a bandwidth of a few GHz is desired.

{
\begin{figure}
\includegraphics[width=1.0\linewidth]{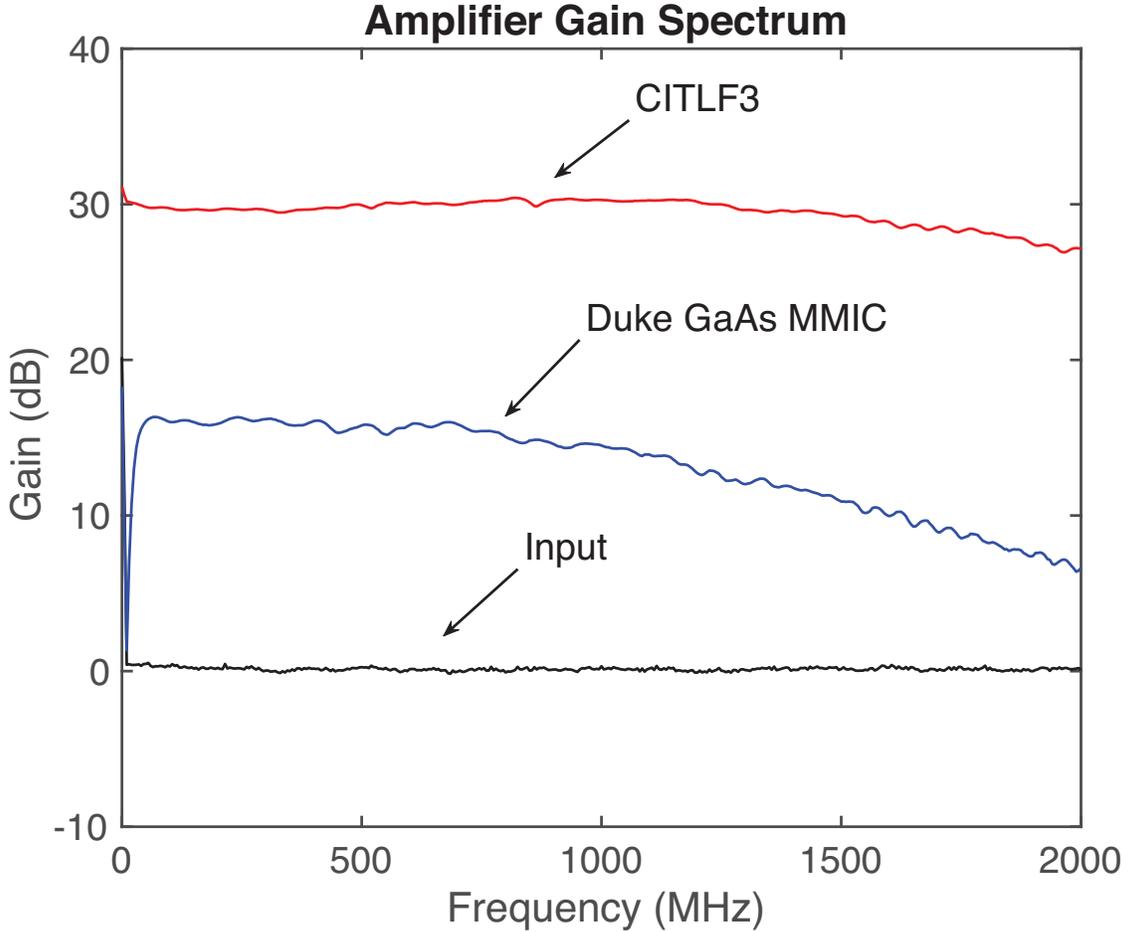}
\caption{Gain spectra are measured using the tracking source of a spectrum analyzer as the input (shown in the black trace for reference). The output of the CITLF\,3 amplifier and the GaAs MMIC circuit are shown in the red and blue traces respectively.}
\label{fig2}
\end{figure}
}

{
\begin{figure}
\includegraphics[width=1.0\linewidth]{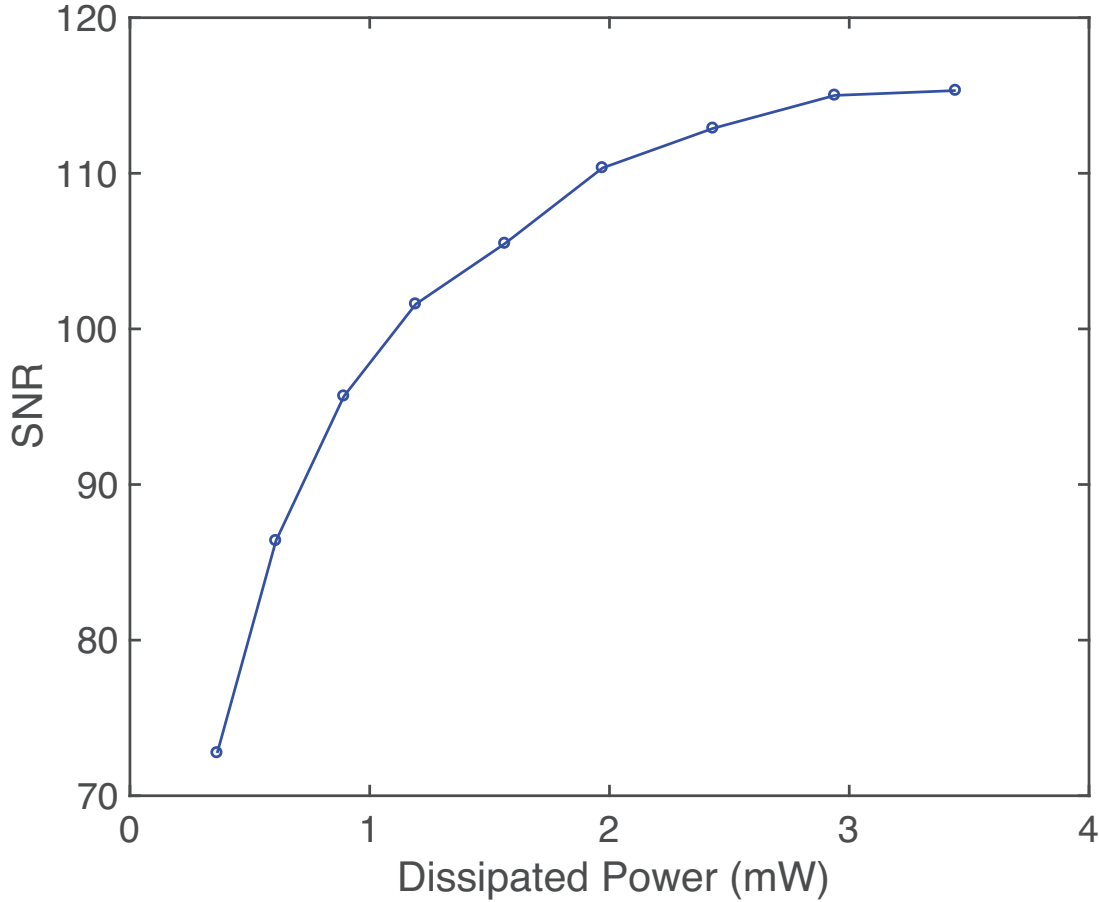}
\caption{SNR of photon detection waveforms recorded at a detector bias current of 12\,$\mu$A as a function of the dissipated power of the GaAs MMIC amplifier.}
\label{fig3}
\end{figure}
}

{
\begin{figure}
\includegraphics[width=1.0\linewidth]{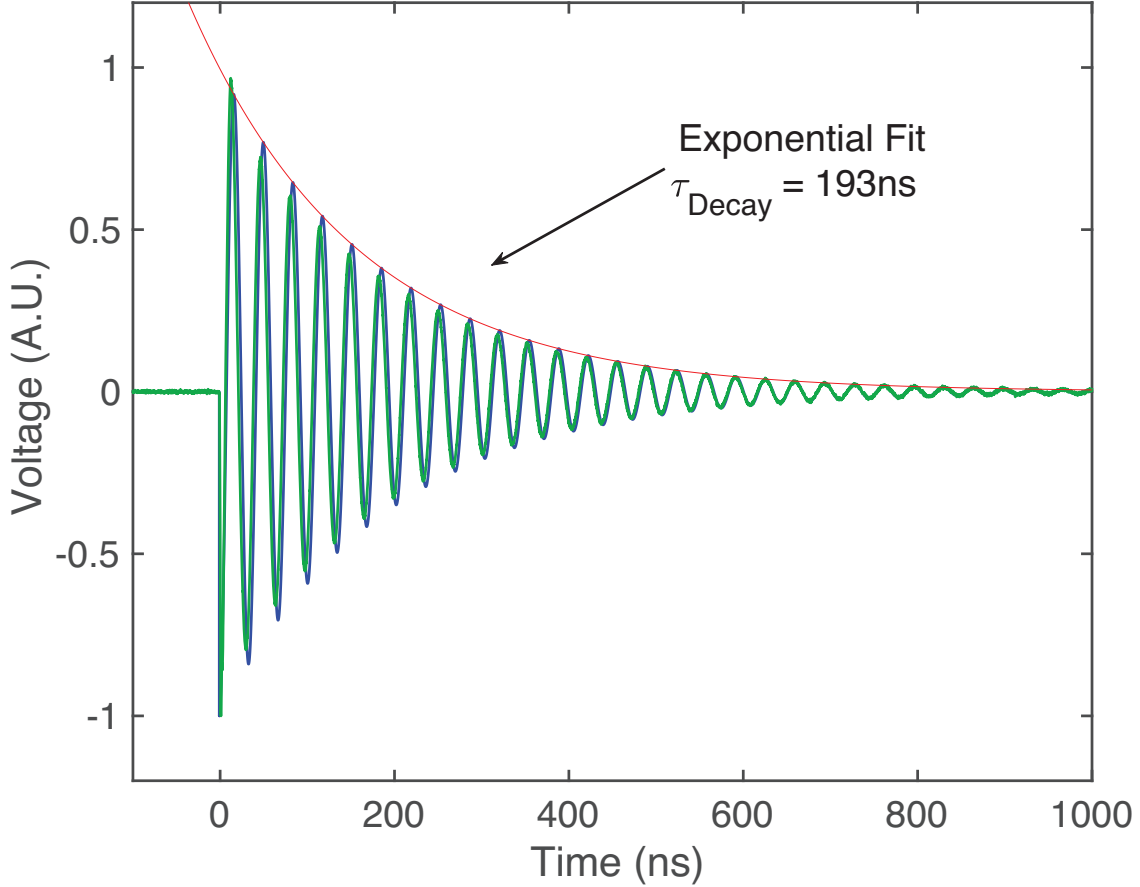}
\caption{Detection event waveform (average of $64$ traces, green) compared to our PSpice circuit model (blue). The decay time is $10$x longer than expected for a $50\,\Omega$ load, indicating that the input resistance of the amplifier chip changes significantly when cooled to $4$\,K.}
\label{fig4}
\end{figure}
}

{
\begin{figure}
\includegraphics[width=1.0\linewidth]{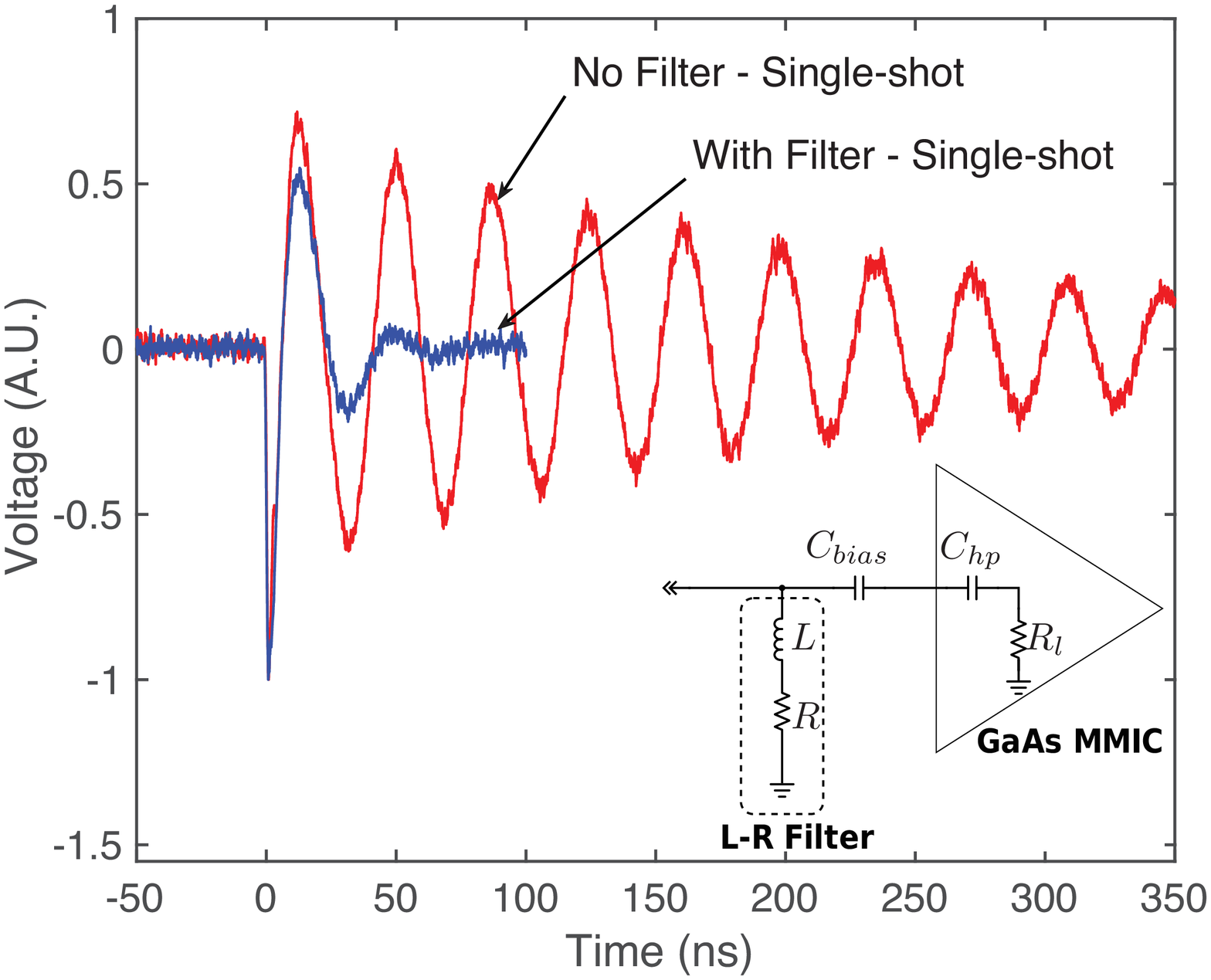}
\caption{Filtered single-shot photon pulse compared to an unfiltered trace averaged pulse. The $1/e$ recovery time $\tau_{Decay}$ is reduced by an order-of-magnitude, while the rise time and SNR are maintained. (\emph{Inset}) Amplifier circuit schematic showing the $L$$-$$R$ filter. The frequency band of the $L$-$R$ path to ground is chosen to include the oscillation frequency. In our circuit $L\sim220$\,nH (Epcos, P/N: B82496C3221J000) and $R\sim100\,\Omega$ (Susumu, P/N: RR1220P-101-D).}
\label{fig5}
\end{figure}
}

Our amplifier has a $1.5$\,GHz bandwidth, gain of $15$\,dB at room temperature, and a $65$\,K noise temperature measured at $4$\,K. The noise temperature is measured at a frequency of 100\,MHz (well within the high-pass band of the $L$-$R$ filter, explained later in this section) using the Y-Factor Method.~\cite{horowitz1989art} The gain spectra of the GaAs MMIC circuit and the CITLF\,3 amplifier, both measured at $300$\,K, are shown in Fig.~\ref{fig2}. One distinct advantage of our circuit is the low power dissipation compared to commercial cryogenic amplifiers. We measure power dissipation of the amplifier operating at a temperature of $4$\,K by recording the supplied current at different bias voltages. As the bias voltage increases, the amplifier gain increases as well as the signal height. The signal-to-noise ratio (SNR), measured at $4$\,K is calculated from the average peak height compared to the variance of the noise floor. The SNR improves as power is dissipated by the amplifier. The SNR levels off at $\sim115$ with $\sim3$\,mW dissipated, shown in Fig.~\ref{fig3}. All subsequent measurements for the GaAs MMIC circuit are performed at this bias point of the amplifier. When compared to the $24$\,mW dissipated by the CITLF\,3 amplifier, an order-of-magnitude reduction in power consumption in our amplifier allows multiple channels to operate simultaneously without overloading the cryogenic system.

The GaAs MMIC chips are not designed to operate at cryogenic temperatures and so their performance drastically changes when cooled to $4$\,K. As shown in Fig.~\ref{fig4}, the waveform observed for photon detection events consists of decaying oscillations with an oscillation frequency of $\sim20$\,MHz and an exponential decay time of $\tau_{Decay}\sim193$\,ns, which is $10\times$ longer than expected for an input load resistance of 50\,$\Omega$. This indicates that the load resistance $R_{l}$ of the amplifier decreases substantially below its room temperature value of 50\,$\Omega$ when it is cooled. To confirm this behavior, we fit the data shown in Fig.~\ref{fig4} to the predictions of a PSpice model where we adjust the value of $R_{l}$ and $C_{hp}$ to match the data. Our model reveals an under-damped LRC resonator, formed by the kinetic inductance of the detector $L_{det}$, the input capacitance of the amplifier $C_{hp}$, and load resistance of the amplifier $R_{l}$. This analysis reveals that $R_{l}\sim5-10\,\Omega$.

The input stage of the amplifier, and the load resistor in particular, is not accessible and so we add an external circuit to modify the effective load impedance and mitigate the slowly decaying oscillations. A slow decay of the photon detection waveform limits the maximum achievable count rate, and therefore a slow decay is unwanted. In order to mitigate this effect and maintain high SNR and maximum bandwidth, we add a $L$-$R$ high-pass filter to the read-out circuit that performs multiple functions. First, it creates a path to ground for all frequencies between DC and its 3-dB cut-off frequency given by $R/2\pi L$. Here, this cut-off frequency is chosen so that the oscillations observed in Fig.~\ref{fig4} are coupled to ground and hence not to the amplifier. An $L$-$R$ filter with the values $L=220$\,nH and $R=100\,\Omega$ is used in the measurements discussed in the following section. Second, it provides a DC path to ground to prevent reverse biasing of the detector at high photon detection rates, as discussed below. Figure~\ref{fig5} shows the photon-detection-event waveform with the filter, where it is seen that the damping time is so short that only a single oscillation cycle is observed. In the context of a quantum communications system with a time-bin encoding~\cite{islam2017provably} the waveform with the filter in place is acceptable for reliably resolving photon detections at a rate of $\sim25\times10^{6}$ counts per second.

\section{Maximum Count Rate}

The maximum achievable count rate (measured in million counts per second, or Mcps) is a performance metric that is heavily influenced by the read-out circuit. The largest contributing factor to the maximum count rate is the input coupling of the first amplifier. As described in Ref.~\onlinecite{kerman2013}, AC-coupled read-out schemes lead to detector saturation at low detection rates due to self reverse biasing of the detector. This effect, due to charge build-up on the input capacitor in the AC-coupled scheme, is present when the detection rate approaches even a small fraction of the detector recovery time. The high-pass filter employed in our setup (Fig.~\ref{fig5}) provides a DC path to ground, avoiding this problem. The room temperature amplifier and CITLF\,3 are AC-coupled, and hence are expected to have degraded saturation characteristics.

We measure the observed count rate as a function of the input optical power with the setup shown in Fig.~\ref{fig6}a. Saturation results in a slower-than-linear count rate as a function of the input power because photons arrive at the detector during the deadtime of the detector. As discussed previously, the deadtime of the detector is determined by $\tau_{Decay}=L_{det}/R_{l}$. The dependence of $\tau_{Decay}$ on the load resistor implies that the read-out circuit will effect the maximum count rate of the detector. The interaction of the detector recovery and the external circuit is discussed in Refs.~\onlinecite{yang2007modeling,kerman2009electrothermal,kerman2013}. We measure the maximum count rate on a single-pixel, amorphous SNSPD with continuous-wave laser light at $1550$\,nm (Wavelength Reference, Clarity-NLL-1550-HP) and a time-tagger (Agilent, Acqiris U1051A). Input flux to the SNSPD is inferred from the measured power reference and a careful calibration of the attenuation in the path to the detector. Our results are shown in Fig.~\ref{fig6}b-c. 

In our GaAs MMIC read-out scheme, the detector operates at a constant bias current of 12\,$\mu$A and the measured counts increase linearly at low input flux until saturation effects become appreciable at the input photon rate of 2 Mcps. The line continues to track smoothly and reaches a maximum measured count rate of 20\,Mcps. This is the point where our time-tagger began dropping significant counts. 

The room temperature amplifier scheme closely follows that observed with the GaAs MMIC circuit until a measured count rate of $\sim10$\,Mcps. To obtain these results with the room temperature amplifier, we had to adjust the detector bias above an input flux of 1\,Mcps to compensate for the current back-action and prevent the detector from latching. The measured counts decrease rapidly at a measured count rate of $\sim10$\,Mcps due to a reduction in the detector bias current to a level such that the efficiency of the detector is decreased substantially.

The detector saturation characteristics using the CITLF\,3 read-out circuit is measured with and without a 3\,dB electrical attenuator at the amplifier input. Without the attenuator, the AC-coupled input to the amplifier causes the detector to latch at a much lower bias current. This latching behavior is worse than that observed with the room temperature amplifier. We believe that the difference between the latching behavior of these two AC-coupled schemes is due to the proximity of the amplifier input to the detector, or to differences in the input circuit between the two amplifiers. Due to the lack of a DC path to ground, the detector bias current had to be adjusted to compensate for latching at each data point. The reduction in bias current is the cause for the poor efficiency. A 3\,dB attenuator installed between the detector and the amplifier provides a DC path-to-ground. In this case, we measure the detector saturation characteristics with a constant detector bias current of 12\,$\mu$A. As shown in Fig.~\ref{fig6}, the 3-dB attenuator substantially improves the counting capabilities of this read-out scheme. The reduction in measured counts when compared to GaAs MMIC and room temperature amplifiers is hypothesized to be the result of a difference in the resistance to ground between our $L$-$R$ filter on our GaAs MMIC circuit and that of the attenuator with the CITLF\,3.

There are other possible front-end circuits that can be placed in front of the CITLF\,3 amplifier to improve its saturation characteristics. For example, a custom $L$-$R$ filter inserted between the amplifier and detector should improve the count rate capabilities of the CITLF\,3 read-out scheme while providing a DC path-to-ground and avoiding signal attenuation as with the 3\,dB attenuator used here. We believe the count rate performance of the CITLF\,3 amplifier with an $L$-$R$ filter could potentially closely follow the saturation characteristics observed for the GaAs MMIC circuit.

{
\begin{figure*}
\includegraphics[width=1.0\linewidth]{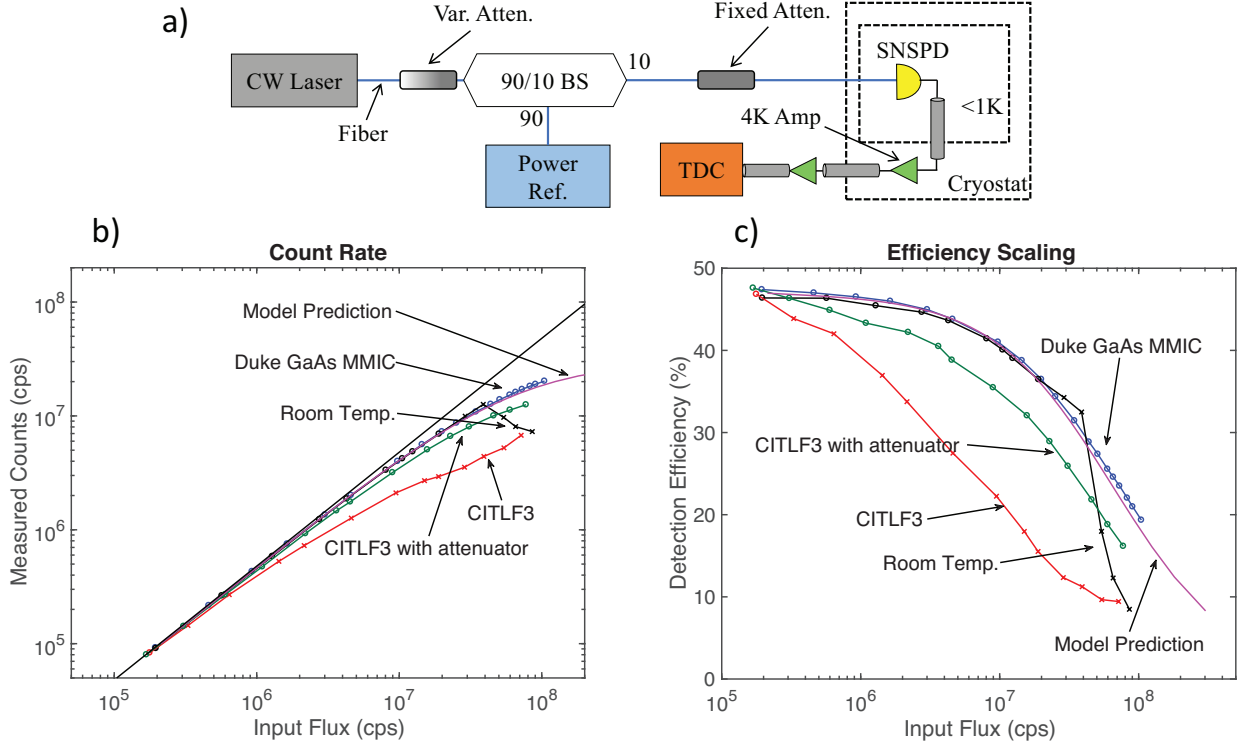}
\caption{\textsf{(a)} Experimental setup for measuring the maximum count rate. \textsf{(b)} Measured count rate as a function of the input optical flux. The black line shown is a fixed efficiency of $48\%$, which is the expected value at $1.55\,\mu$m for this device optimized for operation at a wavelength of $1\,\mu$m. The purple line shows the expected count rate for a 50\,$\Omega$ load resistance as predicted by the model in Ref.~\onlinecite{kerman2013}. The poor performance of the CITLF\,3 without the attenuator is caused by premature latching of the detector due to the amplifier's AC-coupling. Data points with circles are recorded at a 12\,$\mu$A detector bias current. Data points with X's are recorded at a reduced detector bias current due to the detector latching. \textsf{(c)} Count rate data in panel \textsf{(b)} plotted as an effective detection efficiency as a function of the input flux. As the input rate is increased, photons impinge more frequently on the detector before it has fully recovered and do not register a count, leading to a reduction in the detection efficiency.}
\label{fig6}
\end{figure*}
}

\section{Timing Performance}

The maximum rate of communication, especially in time-bin encoded protocols, is affected by the timing resolution of the detection system. We measure the timing resolution of our detector-readout circuit combination using a mode-locked pulsed laser source that emits a train of $\sim5$-ps-width pulses with a repetition rate of $75$\,MHz at a wavelength of $1030$\,nm. A schematic of the measurement setup is shown in Fig.~\ref{fig7}a. The output beam is split by a polarizing beam splitter (PBS), and one beam is measured with a fast photodiode (Miteq DR-125G-A) that has a bandwidth of $12.5$\,GHz and a jitter of $<5$\,ps to serve as a stable timing reference for the pulse arrival. The other beam is heavily attenuated to the single-photon level and the attenuation of the path is adjusted so that the detection rate is $\sim100$\,kcps (mean photon number $\sim10^{-3}$ per pulse). 

A time-correlated, single-photon counting (TCSPC) module (PicoHarp 300 from PicoQuant) measures the time delay between the reference signal and the single-photon signal. The delay due to optical and electrical path length differences is adjusted so that the single-photon event comes after the reference event. We then record a time-of-arrival histogram, where the full-width at half-maximum (FWHM) of a Gaussian fit to the distribution characterizes the timing jitter of the detection system. The timing performance of the GaAs MMIC scheme is compared to both the CITLF\,3 amplifier and low-noise commercial room temperature amplifier. The results of our measurements are shown in Fig.~\ref{fig7}b-c.

Low-noise room temperature amplifiers are the standard read-out scheme for SNSPDs and the timing performance of this scheme is a comparison benchmark for the other read-out schemes. The best timing jitter for room temperature amplifiers is 50\,ps at a detector bias current of 13.1\,$\mu$A. The CITLF\,3 achieves the lowest timing jitter at each detector bias current and reaches a minimum value of 33.5\,ps at a detector bias current of 10.4\,$\mu$A. A higher bias current was not accessible due to the detector latching. Adding the attenuator before the CITLF\,3 enables a higher achievable count rate but it also degrades the photon detection signal and negatively influences the timing jitter. The addition of an $L$-$R$ filter between the CITLF\,3 and the detector in place of the 3\,dB attenuator would provide a DC path-to-ground without degrading the detection signal. We believe the detection jitter of the CITLF\,3 read-out scheme with an $L$-$R$ filter would be better than with the 3\,dB attenuator and should closely follow the measurement without the attenuator. The timing performance of the GaAs MMIC read-out circuit very closely follows the performance of the CITLF\,3 amplifier with the attenuator on the input, and has a minimum timing jitter of 35.4\,ps at a bias current of 13.1\,$\mu$A.

{
\begin{figure*}
\includegraphics[width=1.0\linewidth]{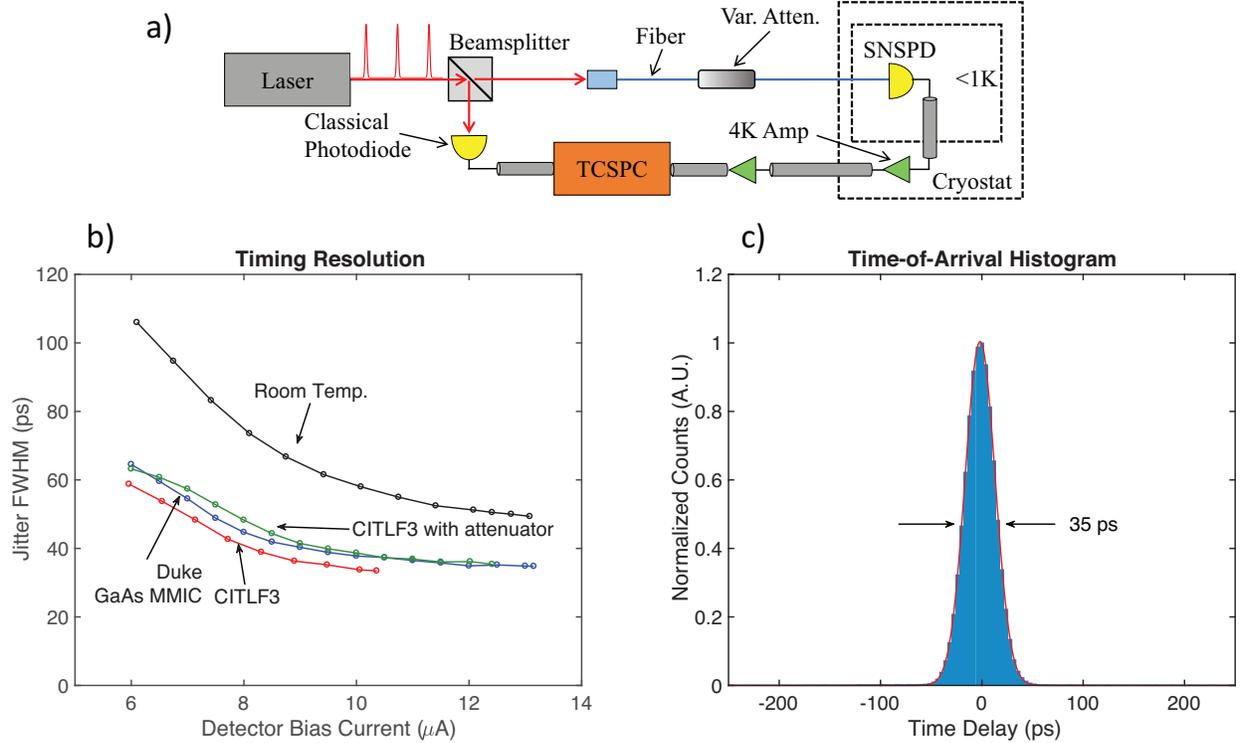}
\caption{\textsf{(a)} Experimental setup for measuring the timing jitter of the photon detection. The time-correlated single-photon counting (TCSPC) module measures the time delay between the reference signal generated by the high-speed classical detector and the single-photon detection signal to build a time-of-arrival histogram. The full-width at half-maximum (FWHM) value of the Gaussian fit to this distribution characterizes the timing resolution of the system. \textsf{(b)} Detection timing jitter as a function of detector bias current for three different read-out schemes - Low-noise room temperature amplifiers (black), our GaAs MMIC circuit (blue), and the CITLF\,3 (red). \textsf{(c)} Time-of-arrival histogram taken using our GaAs MMIC read-out scheme at a bias current of $13.1\,\mu$A. A Gaussian fit to the distribution is shown in the red line.}
\label{fig7}
\end{figure*}
}

\section{Conclusion}

We present a low-cost read-out solution for an SNSPD system that is constructed using commercial GaAs MMIC amplifiers. Our results are highlighted by low noise and high SNR, allowing for high timing resolution while enabling high count rate capabilities. We demonstrate a timing jitter performance as low as $35$\,ps in an amorphous SNSPD, which is a substantial improvement over room temperature amplifiers and comparable to the performance of leading commercial cryogenic amplifiers. Count rates over $20$\,Mcps are observed with our circuit when using a filter before the amplifier that provides a DC path to ground. Our circuit has nearly a factor of $10$ less power consumption compared to leading commercial options, which allows a scalable multi-channel read-out scheme to be realized. 

\section*{Acknowledgements}
This work is supported by Office of Naval Research (ONR) grant on Wavelength-Agile QKD in a Marine Environment (N00014-13-1-0627), and National Aeronautics and Space Administration (NASA) grant on Superdense Teleportation (NNX13AP35A). We thank Drs. Aaron Miller and Tim Rambo for providing SNSPDs for our experiment and helpful discussions. We also thank Vinay Gowda and the Prof. David Smith research group at Duke University for assistance in PCB etching.

\section*{Author contributions}
C.C. designed and built the amplifiers, performed the experiments, and wrote the manuscript. D.J.G. suggested design changes to the read-out circuit, helped interpret the data, and helped write the manuscript. J.K. suggested the GaAs MMIC devices for the experiment, advised various improvements and interpretation of the data, and reviewed the manuscript.

\bibliography{biblist}

\end{document}